\pdfoutput=1
\documentclass[%
 reprint,
superscriptaddress,
 amsmath,amssymb,
 aps,
floatfix,
]{revtex4-2}

\usepackage{graphicx}
\usepackage{dcolumn}
\usepackage[T1]{fontenc} 
\usepackage{siunitx}
\usepackage{verbatim}
\DeclareSIUnit{\belmilliwatt}{Bm}
\DeclareSIUnit{\e}{e}
\DeclareSIUnit{\dBm}{\deci\belmilliwatt}
\usepackage{amssymb}
\usepackage{xcolor}
\usepackage{braket}
\usepackage{bm}
\usepackage{booktabs}
\usepackage{hyperref}

\begin{document}
\raggedbottom

\preprint{APS/123-QED}

\title{A direct dispersive signature of Pauli spin blockade}

\author{Simon~Svab}
\email{e-mail: simon.svab@unibas.ch, dominik.zumbuhl@unibas.ch}
\affiliation{Department of Physics, University of Basel, Klingelbergstrasse 82, \\ CH-4056 Basel, Switzerland}

\author{Rafael~S.~Eggli}
\thanks{This author contributed equally to this work as the first author}
\affiliation{Department of Physics, University of Basel, Klingelbergstrasse 82, \\ CH-4056 Basel, Switzerland}

\author{Taras~Patlatiuk}
\affiliation{Department of Physics, University of Basel, Klingelbergstrasse 82, \\ CH-4056 Basel, Switzerland}

\author{Miguel~J.~Carballido}
\affiliation{Department of Physics, University of Basel, Klingelbergstrasse 82, \\ CH-4056 Basel, Switzerland}

\author{Pierre~Chevalier~Kwon}
\affiliation{Department of Physics, University of Basel, Klingelbergstrasse 82, \\ CH-4056 Basel, Switzerland}

\author{Dominique~A.~Tr\"ussel}
\affiliation{Department of Physics, University of Basel, Klingelbergstrasse 82, \\ CH-4056 Basel, Switzerland}

\author{Ang~Li}
\affiliation{Department of Applied Physics, TU Eindhoven, Den Dolech 2, 5612 AZ Eindhoven, The Netherlands}

\author{Erik~P.~A.~M.~Bakkers}
\affiliation{Department of Applied Physics, TU Eindhoven, Den Dolech 2, 5612 AZ Eindhoven, The Netherlands}

\author{Andreas V. Kuhlmann}
\affiliation{Department of Physics, University of Basel, Klingelbergstrasse 82, \\ CH-4056 Basel, Switzerland}

\author{Dominik~M.~Zumb\"uhl}
\email{e-mail: simon.svab@unibas.ch, dominik.zumbuhl@unibas.ch}
\affiliation{Department of Physics, University of Basel, Klingelbergstrasse 82, \\ CH-4056 Basel, Switzerland}

\begin{abstract}

Pauli Spin Blockade (PSB) is a key paradigm in semiconductor nanostructures and gives access to the spin physics. We report the direct observation of PSB with gate-dispersive reflectometry on double quantum dots with source-drain bias. The reservoir charge transitions are strongly modulated, turning on and off when entering and leaving the blockaded region, consistent with a simple model. Seen with holes in Ge and Si, the effects are enhanced with larger bias voltage and suppressed by magnetic field. This work lays the foundation for fast probing of spin physics and minimally invasive spin readout. 

\end{abstract}

\maketitle

\emph{Introduction} The exclusion principle put forward by Pauli \cite{Pauli1925} is a fundamental signature of quantum mechanics, ensuring proper symmetry of the wavefunction under particle exchange. In semiconductor nanostructures, the principle manifests as Pauli spin blockade (PSB) for fermions such as electrons \cite{Ono2002} or holes \cite{Li2015}. PSB is both a fundamental paradigm of condensed matter quantum statistics \cite{Ashcroft76} and a key mechanism that has enabled high-fidelity spin readout in quantum dot systems. Since its first observation in the form of current rectification in a double quantum dot (DQD), caused by spin-selective tunneling due to the Pauli exclusion principle \cite{Ono2002}, the field of spin-based quantum computing \cite{Loss1998} has progressed greatly. The compatibility with industrial semiconductor
manufacturing processing has promoted promising scaling of qubit numbers in recent years \cite{Hendrickx2021, Philips2022, Weinstein2023, Neyens2023}, and isotopic purification of silicon has led to the realization of high-fidelity two-qubit gates \cite{Tanttu2023}.

These experimental demonstrations have crucially relied on PSB as an efficient mechanism for spin readout. 
Furthermore, PSB can be leveraged for qubit initialization, facilitating the operation and readout of qubits at elevated temperatures above 1 K \cite{Yang2020, Petit2020, Huang2023, Carballido2024}, and even beyond liquid helium temperatures \cite{Camenzind2022}.
While PSB provides the mapping of the quantum dot spin state to a more easily detectable charge state, there still remains the challenge of measuring this state with high fidelity and speed. 

An attractive method for direct in-situ charge sensing is gate-dispersive reflectometry \cite{Colless2013,GonzalezZalba2015,West2019, Urdampilleta2019, Crippa2019, Lundberg2020}. This requires no additional on-chip elements beyond the quantum dot gates and is also less invasive than sensing with an adjacent quantum dot, since sensor tunneling induces backaction  \cite{Ferguson2023}. Using gate reflectometry, the quantum capacitance of a given charge state can be probed at the interdot transition, providing information on the energy level diagram as it is proportional to the curvature of the energy level with respect to the probe gate voltage \cite{House2015, GonzalezZalba2015}. This allows to resolve anticrossings in a DQD energy level diagram originating from tunnel coupling and other off-diagonal terms in the Hamiltonian of the system. However, directly detecting PSB in gate-sensing experiments can prove challenging, as large magnetic fields are required and non-reciprocities are encountered depending on the detailed system parameters \cite{Lundberg2024}. Furthermore, strong spin-orbit interaction and highly variable site-dependent and anisotropic g-tensors, as present in hole spin systems in Si \cite{Liles2021, Geyer2022} and Ge \cite{Jirovec2022, Froning2021a}, complicate the two-spin level diagram even more, resulting in highly variable dispersive signatures of PSB.

In this Letter, we present the dispersive signatures of the PSB using radio frequency (RF) reflectometry at finite bias voltage and compare it to the DC current. The dispersive signal of the reservoir transitions is strongly modulated by the PSB: at the upper triple point, the dispersive signal is suppressed in the leaking and enhanced in the blockaded regime, strongly alternating its magnitude. At the other triple point, this pattern is reversed, while the DC transport looks very similar at both triple points. In a magnetic field, we observe the simultaneous lifting of PSB both in the DC transport and in the gate-dispersive signal.
These results provide insight into the intricate spin physics and open an alternative route for gate-dispersive spin qubit readout at the reservoir transitions.


\begin{figure}[htb]
\centering
\includegraphics[width=1\linewidth]{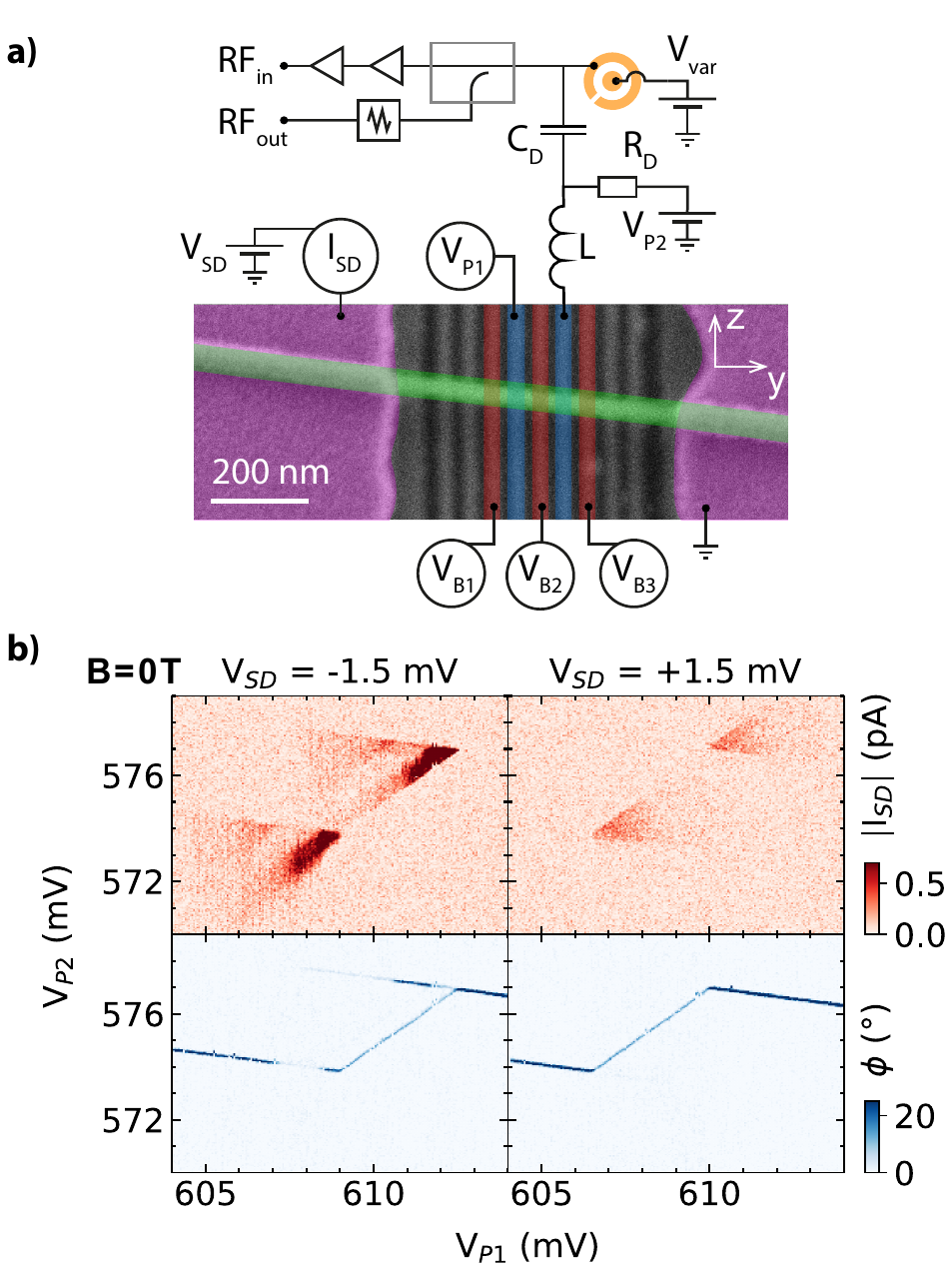} 
\caption{\textbf{Measuring the Ge/Si NW device}. (a) False-colored scanning electron micrograph of a Ge/Si NW device similar to the one used in the experiment (NW angle and curvature may slightly differ). The reflectometry is connected to the right plunger gate for dispersive sensing, impedance matched with the varactor shown in orange, see \cite{EggliSvab2023} for details.
\mbox{(b) Measured} DC signal $\vert I_{\rm{SD}}\vert$ (top) and reflected phase $\phi$ (bottom) in a DQD gate configuration, shown for $V_{\rm{SD}} = \pm 1.5~\rm{mV}$ and $B= 0~\rm{T}$. At $V_{\rm{SD}} = -1.5~\rm{mV}$, the pair of bias triangles exhibits the DC signature of PSB (see Supplemental Material).}
\label{fig:overviewfeature1}
\end{figure}

\ \\

\emph{Dispersive signature of PSB} The experimental setup, including the Ge/Si core/shell nanowire (NW) device \cite{Xiang2006, Hao2010, ConesaBoj2017, Froning2018} used in this work, is depicted in Fig.  \ref{fig:overviewfeature1}a. More details on the device and setup can be found in Ref. \cite{EggliSvab2023}. These NWs can host highly tunable and repeatable hole quantum dots \cite{Higginbotham2014a, Roddaro2008, Brauns2016, Zarassi2017}, and the strong, electrically tunable direct Rashba spin-orbit interaction arising from the heavy-hole light-hole mixing \cite{Kloeffel2011, Kloeffel2018} can be harnessed to achieve fast and all-electrical spin control which is compatible with dispersive gate sensing \cite{Eggli2024}. We furthermore observe high gate lever arms of up to 0.4 in these devices, which facilitates charge sensing through device \mbox{gates \cite{Ungerer2022, EggliSvab2023}}.

We form a DQD by applying positive voltages to five of the nine bottom gates (indicated in Fig. \ref{fig:overviewfeature1}a), locally depleting the hole gas in the NW. The barrier gates (red, at $V_{B1,B2,B3}$) are operated at more positive voltages, creating appropriate tunneling rates, while the plunger gates (blue, at $V_{P1,P2}$) control the hole occupation. The right plunger gate is connected to further circuitry including a surface-mount ceramic core inductor, a bias tee and a voltage-tunable capacitor made of strontium titanate \cite{EggliSvab2023, Apostolidis2020}. The NW is contacted on both ends, allowing for the simultaneous measurements of the direct current $I_{\rm{SD}}$ and the gate-dispersive signal, which is demodulated to give an amplitude and a phase component. Throughout this letter, we define the reflected phase away from the charge transitions as $\phi =$ \SI{0}{\degree} and display the modulus phase change $\phi$.

\begin{figure}[t]
\centering
\includegraphics[width=1\linewidth]{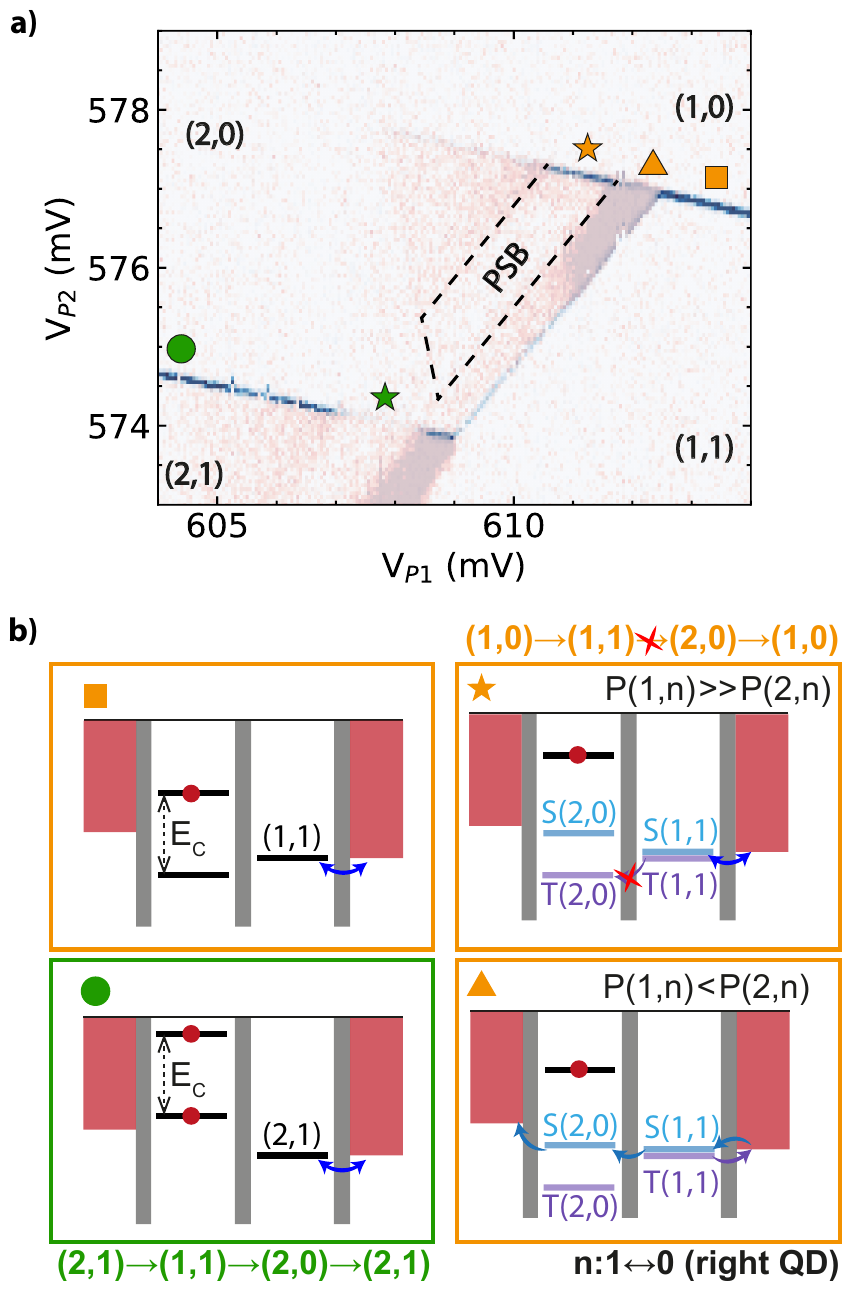} 
\caption{\textbf{Dispersive signature of PSB}. (a) Close-up of the scan at $V_{\rm{SD}} = - 1.5~\rm{mV}$, with superimposed $\vert I_{\rm{SD}}\vert$ (red color scale) and $\phi$ (blue color scale). The region between the dashed lines demarcates the PSB area with suppressed current in the upper triangle. The two lead transitions show complementary but opposite dispersive signatures between the upper and lower lead transitions, see star and triangle regions. This is consistent with the PSB mechanism affecting the transport cycles, see main text. (b) Schematic illustration of the DQD energy levels and the relevant transport cycles shown for different points along the lead transitions, as indicated in (a). In a band close to zero detuning (orange triangle), PSB is lifted and a weak leakage current flows, presumably due to spin-flip cotunneling.}
\label{fig:overviewfeature2}
\end{figure}

At a finite source-drain bias voltage $V_{\rm{SD}}$, scanning the plunger gates $V_{\rm{P1}}$ and $V_{\rm{P2}}$ against each other and measuring the DC signal reveals bias triangles, the characteristic signature of a DQD, as shown in the upper panels of Fig. \ref{fig:overviewfeature1}b.
We plot the simultaneously measured phase $\phi$ in the lower panels, and note two different types of transitions: Interdot transitions correspond to tunneling events between the two quantum dots and trace the baseline of a bias triangle, extending diagonally from lower left to upper right. The lead transitions, on the other hand, follow the other flanks of the bias triangles. Here, we see only the upper flanks (nearly horizontal transitions), since the right plunger gate $P2$ (right dot) is connected to the reflectometry. This detects the tunneling between the right quantum dot and its adjacent reservoir. The direct correspondence of $I_{\rm{SD}}$ and $\phi$ is confirmed in Fig. \ref{fig:overviewfeature2}a, where the data are superimposed. 

We observe the characteristic DC signature of PSB in the presence of strong spin-orbit mixing \cite{Danon2009, Froning2021_soi}: 
At negative bias voltages $V_{\rm{SD}}$ in this case, $\vert I_{\rm{SD}} \vert$ increases in the base region of the triangle when the applied magnetic field is increased, see Supplemental Material \ref{chap:bfield}. This is due to coupling of the spin-blockaded triplet states to the singlet state when the magnetic field is turned on, thereby lifting the spin blockade. At positive $V_{SD}$, we observe no such dependence on the magnetic field, as expected, since PSB is visible only for one bias direction \cite{Ono2002}.

Here, it is worth pointing out a few subtleties of the DC data shown in Figs. \ref{fig:overviewfeature1}b and \ref{fig:overviewfeature2}a. Inside the upper triangle, and along the upper lead transition for negative bias, there is a band of weak current. We assign this effect to reservoir exchange partially lifting the blockade \cite{Johnson2005,Shaji2008}. Another effect is a band of current starting from the zero-detuning line inside the triangle. The size of this band changes with the interdot tunneling strength (see Supplemental Material \ref{chap:cotunneling}) and is caused by some mechanism lifting the PSB, presumably due to spin-flip cotunneling \cite{Qassemi2009, Lai2011, Coish2011}. In this band, the current increases when approaching the upper flank of the bias triangles, consistent with cotunneling \cite{Lai2011}, and suggesting asymmetric reservoir tunneling rates, here with a stronger coupling to the right reservoir. The width of this feature could be related to the exchange splitting $J$. This appears in numerous places in the current, see Ref.~\cite{Golovach2004}, where bias effects were considered, albeit not along the detuning axis. Finally, we note that there are lines, parallel to the left lead transitions (tracing the other flank of the bias triangles), at which the current changes (this is even more pronounced for a larger bias window, see Supplemental Figs. \ref{fig:psb_dcscans}a and \ref{fig:psb_dcscans}b). This can be explained with (ground and excited) states of the left QD being aligned with the Fermi level of the left reservoir \cite{Hanson2007}.

Next, we take a closer look at the lead transitions in reflectometry. Each QD contains dozens of holes, but for the case of PSB the effective charge states correspond to those indicated in Fig. \ref{fig:overviewfeature2}a. The energy level diagrams for some representative configurations (indicated with colored markers) are drawn in Fig. \ref{fig:overviewfeature2}b. The (1,0)$\leftrightarrow$(1,1) lead transition (orange square) corresponds to the (1,1) energy level of the right QD being aligned with the Fermi level of the right reservoir. This leads to resonant tunneling with the reservoir (blue arrow) and a change of the right QD charge state as this line is crossed. The corresponding change of the dot level population appears in the tunneling capacitance and is picked up by the reflectometry \cite{Vigneau2022}.
When the energy levels of the left QD lie outside the bias window, there is no DC transport through the DQD due to Coulomb blockade, while a finite $\phi$ can be measured on the lead transition.

Inside the dashed lines in Fig. \ref{fig:overviewfeature2}a, PSB causes the DQD to remain stuck in the T(1,1) state for most of the time, such that the measured $\vert I_{\rm{SD}} \vert$ is very low. Transport can occur at a large enough detuning where the T(2,0) state becomes energetically accessible. At low detuning, PSB is lifted by some process such as cotunneling, providing a band of current from the zero detuning up to the dashed line. Finally, there can also be reservoir exchange along the upper edge of the bias triangle. In presence of PSB leakage, the resonant tunneling with the right reservoir is suppressed due to the slow left reservoir tunneling. This keeps the right dot in the 0 state for most of the time and thus the change of population across the reservoir line is weak, as seen by the almost completely suppressed gap in the dispersive sensing at the orange triangle. In contrast, the dispersive signal is high in the spin-blockaded region (orange star), although not quite as strong as in the Coulomb blockade region (orange square). To check the important role of the asymmetric DQD setup as hypothesized above, we increase the left dot tunneling rate and observe that the dispersive signal indeed recovers in strength, as expected, see Supplemental Fig.~\ref{fig:leftwall}.

Shifting the focus to the (2,0)$\leftrightarrow$(2,1) lead transition (green circle), this is the case of two holes on the left dot, while the energy level of the right QD is still aligned with the Fermi level of the right reservoir. In analogy to the orange square region, but now with two holes on the left dot, the (2,0)$\leftrightarrow$(2,1) transition produces a change in the charge state of the right QD, leading to a change in the tunneling capacitance. We note that this transition traces the other bias triangle, whose transport cycle is different -- namely, the occupation in the DQD is cycling between 2$\leftrightarrow $3 holes instead of 1$\leftrightarrow$2, as labeled in Fig. \ref{fig:overviewfeature2}b. Nonetheless, the transport cycle involves an occupation where the DQD can get stuck in the T(1,1) state due to PSB, leading to a reduced $\vert I_{\rm{SD}} \vert$ and a suppression of right reservoir tunneling, thus turning off the tunneling capacitance and phase signal, see the gap in the lower reservoir transition (green star). 

We note that the dispersive signals at the edges of the two triangles are complementary and opposite in behavior: for example at the starred regions (orange and green stars), the PSB is mostly intact and keeps the DQD blocked in the T(1,1) state for most of the time. At the upper triangle (orange star), the reservoir transition (1,1)$\leftrightarrow$(1,0) remains active with resonant tunneling with the reservoir ongoing, producing a strong dispersive signal. On the other hand, at the lower transition (green star), the occupation of the T(1,1) in PSB mostly deactivates the right reservoir transition (2,1)$\leftrightarrow$(2,0), giving a weak dispersive signal, as indeed seen in the data. 

Similar signatures of the dispersive PSB were also observed in a Si FinFET hole spin qubit device, now with absolute hole numbers one/two, see Supplemental Material \ref{supplement_Finfet}. The dispersive gap at the upper reservoir transition is present as well as the reappearance of the dispersive signal when PSB is intact, and its mirror image at the lower reservoir. The dispersive gap is maybe slighly weaker or smaller, but this also depends on the interdot tunneling (see Fig. \ref{fig:interdot}) as well as the left dot rate (see supplement \ref{chap:leftwall}). Thus, this confirms the same key signatures also in a different material and device type. 

We now aim to provide a phenomenological explanation to our observation. The measured change in $\phi$ is proportional to the parametric capacitance $C_{\rm{PM}} = C_Q + C_{\rm{TU}}$, where $C_Q$ is the quantum capacitance, due to the curvature of the (occupied) energy levels, i.e. from the singlet-triplet anticrossing. The tunneling capacitance $C_{\rm{TU}}$ \cite{Betz2015} can be written as \cite{Mizuta2017, Lundberg2024}:

\begin{equation}
C_{\rm{TU}} = (e\alpha)^2 \sum_{i} \left(\frac{1}{2}\left(1 + 2\frac{\partial E_i}{\partial \varepsilon} \right) \frac{\partial P_i}{\partial \varepsilon}\right) ~~,
\label{tcapacitance}
\end{equation}

\noindent where $\alpha$ is the gate lever arm, $P_i$ the occupation probability of each individual state with energy $E_i$, and $\varepsilon$ is the DQD detuning. This equation takes into account the change of the dot occupations and the slope of the energy levels, and considers only the internal DQD transition. A similar term applies to the lead transitions. The occupations change in a step like manner at the charge transitions, where a large $C_{\rm{TU}}$ results for the interdot as well as the lead transitions. Further, PSB influences the populations, in particular leading to a reappearance of the reservoir line at the orange star in the blockaded regime, and conversely a suppression at the green star in Fig.~\ref{fig:overviewfeature2}a. The quantum capacitance is visible near the zero-detuning in our data set, where the curvature of the DQD anticrossing appears, and gives no significant contribution to the lead transitions. 

\ \\

\begin{figure}[t]
\includegraphics[width=1\linewidth]{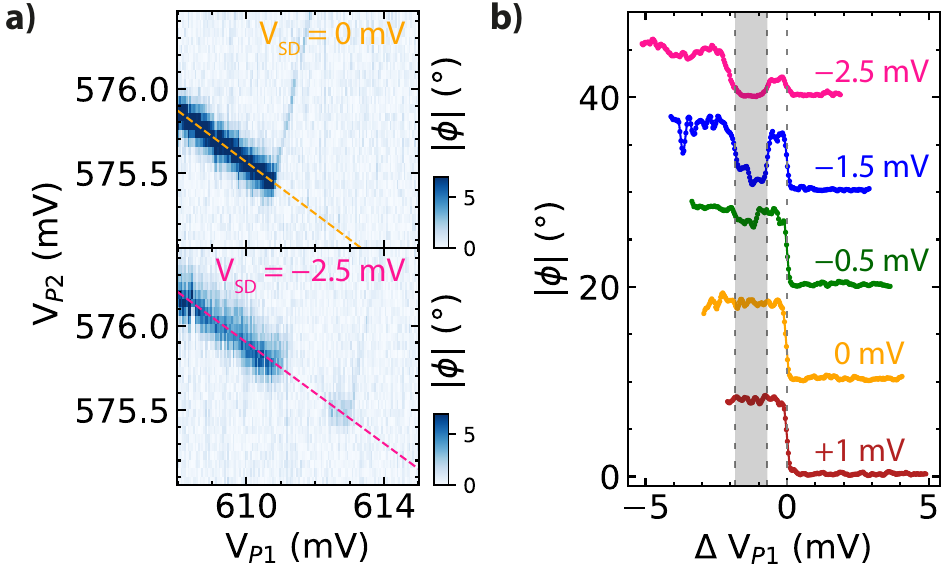} 
\caption{\textbf{Bias dependence}. (a) Dispersive phase shift $\vert\phi\vert$ at the (2,0)$\leftrightarrow$(2,1) lead transition, shown at $B= 0~\rm{T}$. Colored dashed lines indicate the line cuts shown in (b) along the lead transition for bias voltages as labeled. These line cuts are obtained by interpolation and smoothing. The zero detuning flank defines the zero of the axis. The gray shading demarcates the region where the DQD is mostly in T(1,1). As $V_{\rm{SD}}$ becomes more negative, the suppression of $\vert\phi\vert$ due to PSB becomes visible (gray area) and also decreases in the cotunneling region, from the edge of the gray band to zero detuning.}
\label{fig:biasdependence}
\end{figure}

\emph{Dependence on experimental parameters} We further investigate the conditions under which the described dispersive signature is modulated. In Fig. \ref{fig:biasdependence}a we show a close-up of the (2,0)$\leftrightarrow$(2,1) lead transition, for two different $V_{\rm{SD}}$ values as labeled. While the strength of the lead transition is essentially constant for zero bias, the dip in the lead signal emerges for more negative voltage bias. We perform line cuts along the lead transition, as indicated with colored dashed lines, and obtain the profiles shown in Fig. \ref{fig:biasdependence}b. For negative bias, the dispersive gap (shaded gray) appears in the PSB region, ultimately approaching the background value, corresponding to absence of reservoir tunneling. This is reminiscent of the original observation of PSB \cite{Ono2002} where the blockaded region in direct current appeared only at bias voltages above which the triplet becomes available. Similarly, the phase signal weakens in the region between the gap and zero detuning as the dispersive gap saturates. These effects illustrate how $C_{\rm{TU}}$ can change inside the bias window. 

\begin{figure}[t]
\centering
\includegraphics[width=1\linewidth]{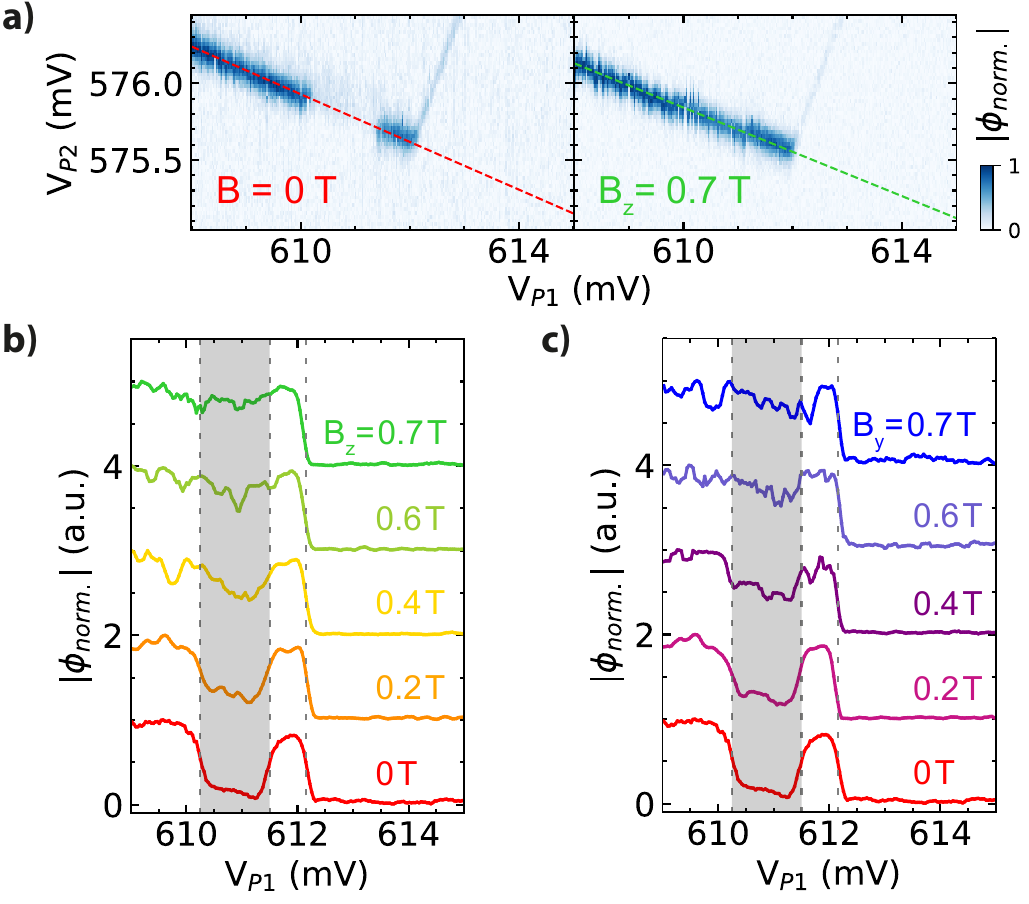} 
\caption{\textbf{B-field dependence}. (a) Normalized phase at the (2,0)$\leftrightarrow$(2,1) lead transition at $V_{SD} = -1.5~\mathrm{mV}$ and magnetic fields as labeled. Colored dashed lines indicate the line cuts shown in (b) and (c), where the magnetic field evolution is shown as labeled. The z-axis is in plane and perpendicular to the NW, and the y-axis is essentially along the NW. With growing field, the visibility of the dispersive gap (gray bar) is restored, similarly for B-field along the y or z-axes. }
\label{fig:bfielddependence}
\end{figure}

Additionally, the dispersive feature depends on the strength of the applied magnetic field. As highlighted in Fig. \ref{fig:bfielddependence}a, the dispersive gap is filled-in and the dispersive signal is restored when the magnetic field is increased.  This is in line with the DC characteristics in this charge configuration (see Supplemental Material \ref{chap:bfield}). Next to the lead transition, one can also see in Fig. \ref{fig:bfielddependence}a that the interdot transition becomes weaker in presence of a B-field, similar to previous reports  \cite{Pakkiam2018, West2019, Crippa2019,Russell2023}. Finally, we note the slightly more noisy data in presence of magnetic field, indicating elevated switching noise, possibly more pronounced for the y-field. This could be due to a somewhat elevated temperature or the B-field shifts of the two level fluctuators.

\ \\

\emph{Conclusions and outlook} We have observed the dispersive signatures of PSB in a DQD at finite bias.
Asymmetric reservoir coupling boosts the visibility of the dispersive signal at the reservoir transition, where the dispersive signals are switching on- and off together with the PSB current. Similar signatures were also observed in a Si FinFET device. A first phenomenological model is consistent with the dispersive PSB signals, but more work is needed for a quantitative understanding, particularly also of the leakage current. This may be complicated by the fact that the DQD is operating with holes and experiences strong spin-orbit coupling. 

The minimally invasive gate reflectometry allows for short integration times, thus providing access to time-resolved spin physics on timescales of $\mu\rm{s}$ and below \cite{House2015, EggliSvab2023}. In future experiments, the outlined dispersive signature of PSB may be harnessed to perform spin readout, thereby adding a new method to the repertoire of existing qubit readout approaches. 
Here, effects that lift PSB such as spin-flip cotunneling could reduce readout fidelity. However, we were able to tune the extent to which this effect was present in our experiment (Supplemental Material \ref{chap:cotunneling}), and that our proposed readout method should work irrespective of it. 
We note that the dispersive sensing applies only a very small (monochromatic) signal of $\mu \rm{V}$ level to one of the gates, thus strongly reducing backaction effects compared to probing with an adjacent quantum dot sensor, which suffers from random electron tunneling fluctuations \cite{Ferguson2023}. 

\ \\

The data supporting this study are available in a Zenodo repository \cite{Zenodo2025}.

\ \\

\emph{Author contributions} S.S., R.S.E. and D.M.Z. conceived of the project and planned the experiments. S.S. fabricated the NW device with contributions from M.J.C. and P.C.K.. S.S. performed the experiments on the Ge/Si NW with inputs from R.S.E. and T.P.. S.S. and R.S.E. outlined the phenomenological explanation of the dispersive signature with inputs from D.M.Z.. R.S.E. performed the experiments with the FinFET device which was fabricated by A.V.K.. T. P. and D. A. T. contributed to the experimental setup. S.S. analyzed the data with inputs from R.S.E., T.P. and D.M.Z.. A.L. grew the NWs under the supervision of E.P.A.M.B.. S.S. wrote the manuscript with inputs from all authors. D.M.Z. supervised the project.

\emph{Acknowledgements} We thank S. Bosco, W. A. Coish, E. R. Franco Diaz and J. C. Egues for useful theory discussions. Further thanks go to A. Efimov and Y. Liu for additional experimental efforts in the Ge/Si NW project, and E. Kelly, A. Orekhov and G. Salis for providing the superconducting nanowire inductors used in the experiments with the FinFET device. Furthermore, we acknowledge S. Martin and M. Steinacher for technical support. This work was partially supported by the Swiss Nanoscience Institute (SNI), the NCCR SPIN, the Georg H. Endress Foundation, Swiss NSF (grant no. 179024), the EU H2020 European Microkelvin Platform EMP (grant no. 824109) and FET TOPSQUAD (grant no. 862046).

\clearpage
\onecolumngrid
\section{Appendix}\label{appendix}

\subsection{Magnetic field dependence of the investigated bias triangles}
\label{chap:bfield}

\begin{figure}[t]
\centering
\includegraphics[width=0.65\linewidth]{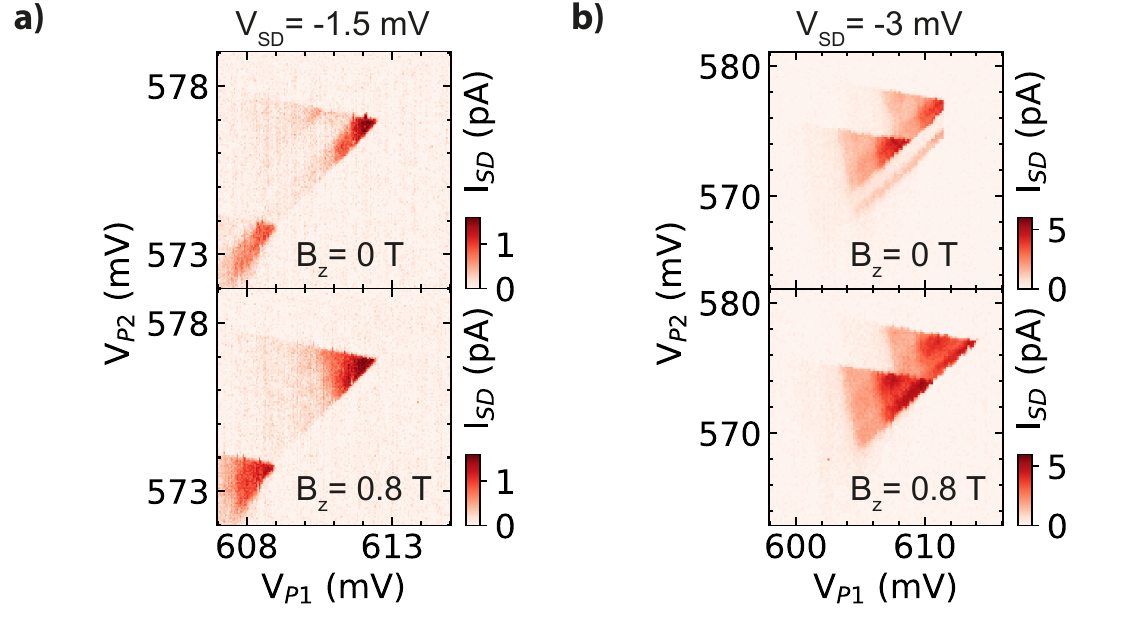} 
\caption{\textbf{Supplementary data on the magnetic field dependence}. Measured DC signal $I_{\rm{SD}}$ at a source-drain bias voltage (a) $V_{\rm{SD}} = -1.5~\rm{mV}$ and (b) $V_{\rm{SD}} = -3~\rm{mV}$. The magnetic field lifts the PSB, resulting in an increase of $I_{\rm{SD}}$ close to the triangle baseline. This data was taken at the same gate configuration as the one presented in the main text.}
\label{fig:psb_dcscans}
\end{figure}

Here, we provide additional information on the data presented in the main text. As a standard configuration, the DQD was operated at barrier gate voltages $V_{\rm{B1}} = 2~\rm{V}$, $V_{\rm{B2}} = 1.82~\rm{V}$ and $V_{\rm{B3}} = 3~\rm{V}$. We note that all scans were performed by ramping from negative to positive voltages, and that $V_{\rm{P1}}$ is the slow axis while $V_{\rm{P2}}$ is the fast axis. Fig. \ref{fig:psb_dcscans}a shows the DC signal closer to the upper bias triangle at $V_{\rm{SD}} = -1.5~\rm{mV}$, for $B= 0~\rm{T}$ (top panel) and  $B_z= 0.8~\rm{T}$ (bottom panel). While not as visible at this low bias voltage, the magnetic field lifts the PSB, such that $I_{\rm{SD}}$ increases in the previously blockaded region. This effect becomes more clear at $V_{\rm{SD}} = -3~\rm{mV}$, as presented in Fig. \ref{fig:psb_dcscans}b.

Interestingly, at this increased bias window size the transport would occasionally get blockaded in the region close to the corner (adjacent to the baseline and the (1,0)$\leftrightarrow$(1,1) lead transition) of the upper triangle at $B= 0 ~\rm{T}$, leading to a seemingly missing section of the bias triangle shown in the upper panel of Fig. \ref{fig:psb_dcscans}b. Scanning $V_{\rm{P2}}$ through this region with a high resolution (i.e. a longer time spent in this region), we note that DC transport through the DQD seemed to get blocked and unblocked every now and then (on the timescale of few tens of seconds). A scan in the same gate configuration (measured several weeks later, after investigating other gate configurations) is presented in Fig. \ref{fig:interdot}a (rightmost panel), showing that this data was well reproducible within the same thermal cycle.

\subsection{Width of the current band near zero detuning}
\label{chap:cotunneling}

\begin{figure}[t]
\centering
\includegraphics[width=1\linewidth]{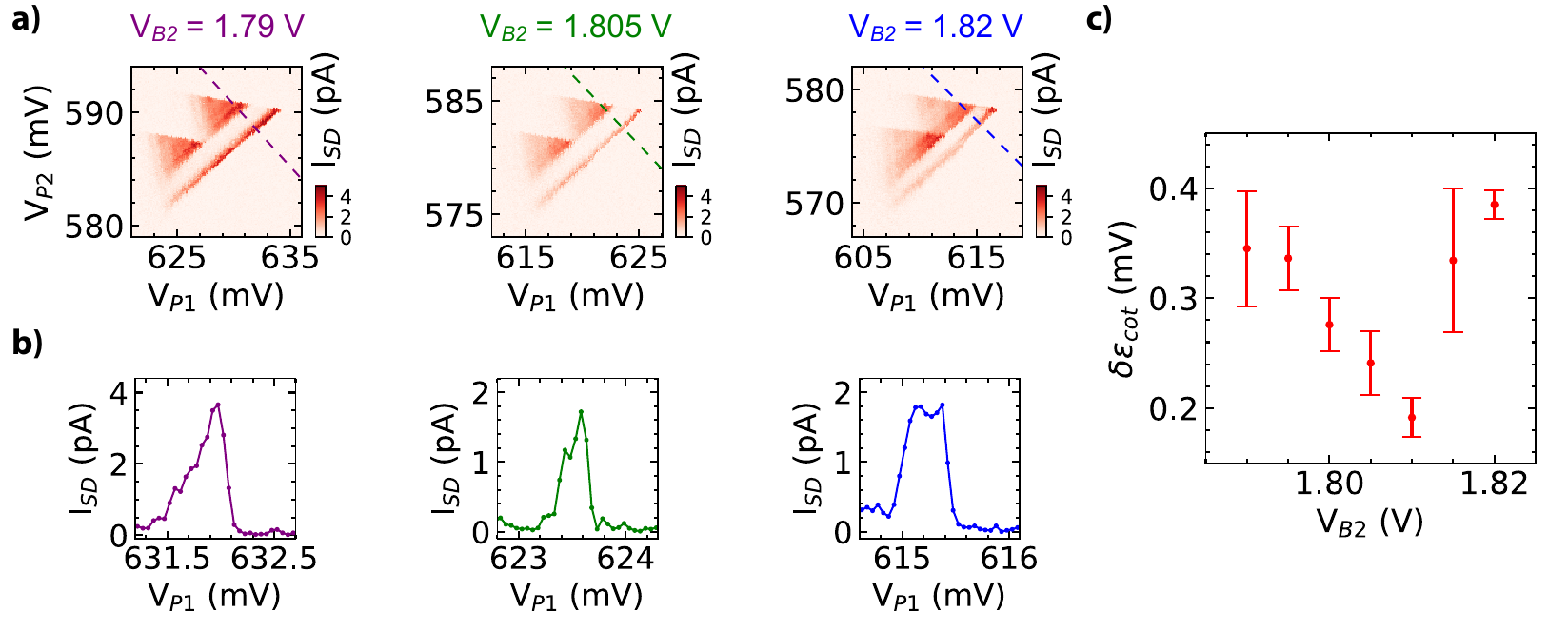} 
\caption{\textbf{Width of the current band}. (a) DC measurements of the investigated pair of bias triangles, shown for different interdot barrier gate voltages $V_{B2}$. (b) Linecuts through the indicated axes in (a), allowing to extract a peak width $\varepsilon_{cot}$. (c) Extracted $\varepsilon_{cot}$ as a function of $V_{B2}$. The error bars correspond to the standard deviation of peak widths obtained from linecuts close to the region of interest.}
\label{fig:interdot}
\end{figure}

In the main text, we found that PSB is lifted in a band close to zero detuning, which may, for example, originate from inelastic (spin-flip) cotunneling \cite{Qassemi2009, Lai2011, Coish2011}. Here, we show that the width of this band depends on the center barrier gate voltage $V_{\rm{B2}}$, and hence the interdot tunneling rate. In Fig. \ref{fig:interdot}a, we present further scans of the bias triangles presented in the main text, taken at different values of $V_{\rm{B2}}$. Qualitatively, it can be seen that the width of the band is more narrow at $V_{\rm{B2}} = 1.805 ~\rm{V}$ than at the two other barrier gate configurations. We confirm this quantitatively by performing nearest neighbor interpolation of the data through the indicated axes, yielding the linecuts as shown in Fig. \ref{fig:interdot}b. In order to better distinguish the narrow peak widths, we do not perform any smoothing on these linecuts. We extract $\delta \varepsilon_{cot}$ as the mean value of the full width at half maximum of $n=10$ close (adjacent measurement points) collinear linecuts in the same triangle regions, and plot the values in Fig. \ref{fig:interdot}c along with the standard deviations as error bars. In our data, we find that $\delta \varepsilon_{cot}$ is halved within this range of $V_{\rm{B2}}$, with a local minimum at $V_{\rm{B2}} = 1.81 ~\rm{V}$.

In the spin-flip cotunneling theory, the dependence of $\delta \varepsilon_{cot}$ on $V_{\rm{B2}}$ could be explained by considering the involved tunneling rates and following Refs. \cite{Qassemi2009, Coish2011, Lai2011}, such that the cotunneling peak width $\delta \varepsilon_{cot}$ at $B=0$ is expressed as
\begin{equation}
    \delta \varepsilon_{cot} = \left( \frac{3\Gamma_S t^2}{W_{cot}^0} \right)^{1/2} ~ .
    \label{cotunnelingpeakwidthequation}
\end{equation}

Here, $W_{cot}^0$ is the spin-flip cotunneling rate, given by
\begin{equation}
    W_{cot}^0 = \frac{k_B T}{\pi \hbar} \left[ \left(\frac{\hbar \Gamma_D}{\Delta - \varepsilon}\right)^2 + \left(\frac{\hbar \Gamma_S}{\Delta + \varepsilon - 2 U_{lr} - 2 \vert e V_{SD} \vert}\right)^2 \right]~,
\end{equation}
where $\Delta$ sets the energy of the (1,1) charge configuration and $U_{lr}$ is the mutual (interdot) charging energy. Taking $\Gamma_S \ll \Gamma_D$ for the source (drain) lead tunneling rates, as is also valid for our data, the expression for the cotunneling rate becomes
\begin{equation}
    W_{cot}^0 \simeq \frac{k_B T}{\pi \hbar} \left( \frac{\hbar \Gamma_D}{\Delta - \varepsilon} \right)^2~,
\end{equation}

and inserting this back into Eq. \ref{cotunnelingpeakwidthequation} yields

\begin{equation}
    \delta \varepsilon \propto \frac{\sqrt{\Gamma_S}}{\Gamma_D} \cdot t~ .
    \label{equation_widthdependence}
\end{equation}

Thus, one can see that $\delta \varepsilon$ is influenced by all three tunneling rates of the DQD system. Circling back to our data in Fig. \ref{fig:interdot}c, and considering that the tunneling rates in our depletion-mode hole DQD decrease as gate voltages become more positive, we argue that $\delta \varepsilon$ decreases from $V_{B2} = 1.79 ~\rm{V}$ to $V_{B2} = 1.81 ~\rm{V}$ as a consequence of the simultaneously decreasing interdot tunneling rate $t$. 

From $V_{B2} = 1.81 ~\rm{V}$ to $V_{B2} = 1.82 ~\rm{V}$, the extracted $\delta \varepsilon$ increases again. Since $t$ should continue to be suppressed more by $V_{B2}$, this hints at the ratio of reservoir tunneling rates $\sqrt{\Gamma_S} / \Gamma_D$ rising in this interval, thereby increasing $\delta \varepsilon$. Indeed, a closer look at the plunger gate voltage values in Fig. \ref{fig:interdot}c shows that the negative compensation of $V_{P1}$ (as $V_{B2}$ becomes more positive) is approximately double that of $V_{P2}$. Thus, we reason that these plunger gates affect their adjacent reservoir tunneling barriers, $\Gamma_S$ and $\Gamma_D$ respectively, and that $\Gamma_S$ is rising more strongly in this gate voltage range, leading to an increase in $\sqrt{\Gamma_S} / \Gamma_D$ and thus a broader $\delta \varepsilon$.

\subsection{Left barrier gate dependence}

\label{chap:leftwall}

\begin{figure}[t]
\centering
\includegraphics[width=1\linewidth]{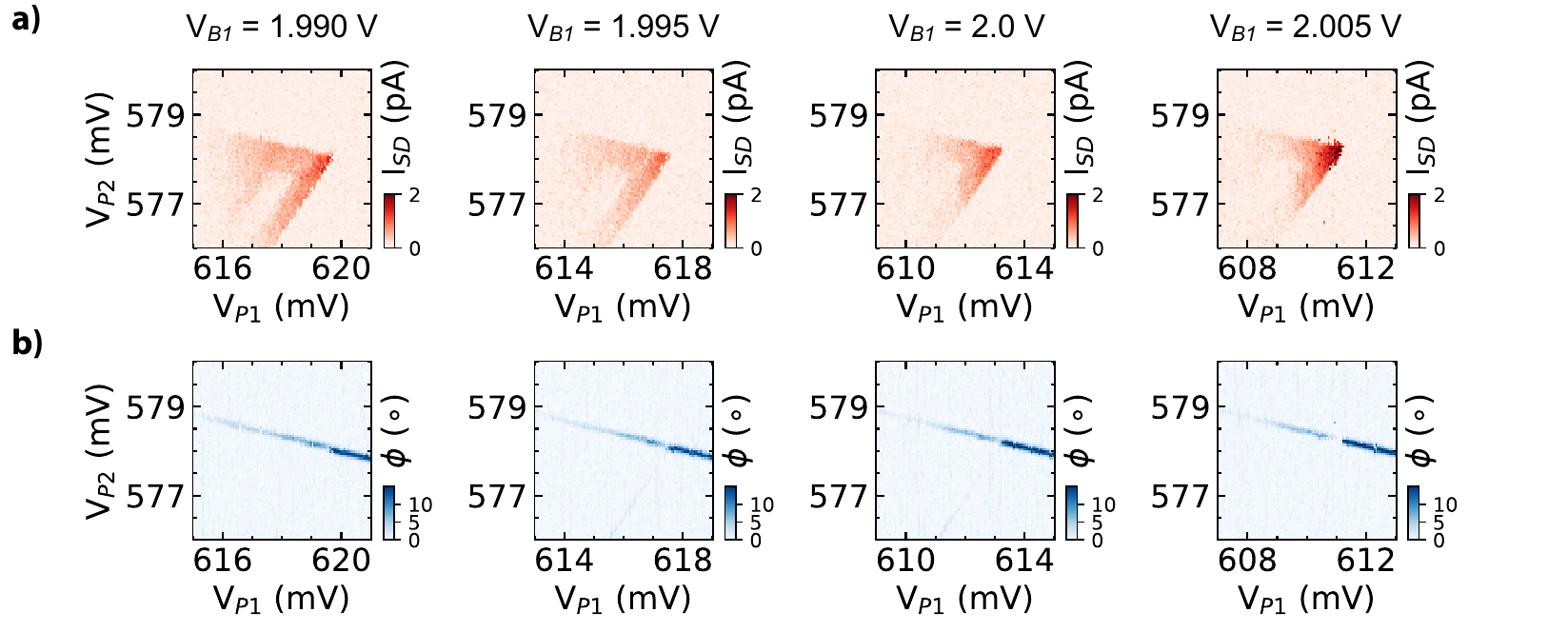} 
\caption{\textbf{Dependence on the left barrier gate voltage}. (a) Measured DC signal and (b) corresponding dispersive signal $\phi$ of the investigated pair of bias triangles, shown for different left barrier gate voltages $V_{B1}$. The gap of the dispersive signal (right panel, lowest left dot tunneling rate) along the reservoir transition closes and the dispersive signal is recovered as the left dot tunneling rate is increased by lowering the barrier $V_{B1}$ (left panel, highest left dot rate).}
\label{fig:leftwall}
\end{figure}

In our experiment, we have also observed that the modulation strength of the sensing signal along the bias triangle changes as one tunes the left barrier gate voltage $V_{B1}$ (corresponding to the source tunneling rate $\Gamma_S$). This can be seen in the data set in Fig. \ref{fig:leftwall}, taken in the same bias triangle as the one shown in the main text, albeit at a later point in time after retuning the device to this voltage configuration. As $V_{B1}$ is increased, we see that the change in contrast (of the dispersive signal) along the lead transition becomes pronounced. As we argue in the main text, this may stem from changes in the transport cycle, and thus larger differences in $\partial P_i / \partial \epsilon$ in the different sectors of the bias triangle. 

\subsection{Additional data from a Si FinFET DQD}
\label{supplement_Finfet}
\begin{figure}[htb!]
\centering
\includegraphics[width=1\linewidth]{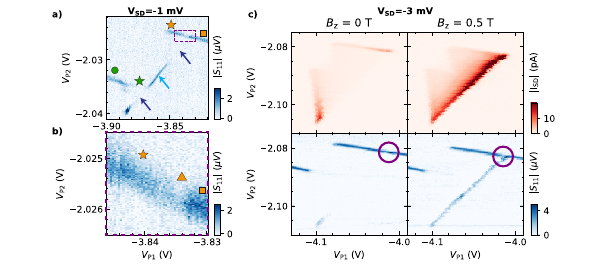} 
\caption{\textbf{Additional PSB data measured with a Si FinFET qubit device}. (a) A dispersively sensed charge stability map (showing the change in reflected amplitude $\vert S_\mathrm{11}\vert$) of the FinFET device \cite{Eggli2024} near the $(2,0)\leftrightarrow(1,1)$, as labeled, transition is shown with the different regimes indicated by yellow and green markers following the definition in Fig \ref{fig:overviewfeature2}. At small bias, the two triple points are sufficiently separated to not overlap. The zero-detuning line (arrows) is visible outside the bias window (light blue arrow) but disappears within the bias triangles (dark blue arrows). As for the NW device, the dispersive PSB signature appears as a bright upper reservoir transition (yellow star) and an invisible lower reservoir transition (green star). Additionally, PSB is only weakly lifted close to zero-detuning, leading to a minute gap in the upper reservoir transition (dashed purple rectangle). A higher resolution scan of this region is shown in (b). The gap in the reservoir transition (orange triangle) is indicated as well as the spin blockade (orange star) and the reservoir transition outside of the bias window (orange square). (c) The DC current (upper panels) through the FinFET shows the characteristic signature of PSB, as the current increases close to the triangle baseline when a magnetic field is applied (right panels). Unlike the Ge/Si NW DQD, the blockade remains largely intact in the inelastic cotunneling regime. Therefore, the reservoir transitions in the dispersive measurement (lower panels) remain largely unaffected apart from a slight increase of the gap in the upper reservoir transition close to zero detuning (purple circles).   }
\label{fig:Supp_Finfet}
\end{figure}

We have found similar dispersive PSB features in a Si FinFET device \cite{Geyer2022,Camenzind2022,Geyer2021} which was cooled down in the same dilution refrigerator but using a superconducting NbN nanowire inductor \cite{Kelly2023, Eggli2024} in the tank circuit. The tank was attached to the right plunger gate. For a detailed description of the device, see Ref. \cite{Eggli2024}. Note that the identical device and charge transition was used in previous works \cite{Eggli2024,Bosco2023} to perform spin qubit experiments using DC-based spin readout \cite{Camenzind2022}. From the absence of further transitions in larger-scale charge stability maps (not shown) we conclude that the DQDs absolute charge occupation was $(1,1)\leftrightarrow(2,0)$. The correspondence of the data taken with the FinFET to the results in the main text is highlighted in Fig. \ref{fig:Supp_Finfet}. The $(1,1)\leftrightarrow(2,0)$ charge transition at relatively small bias $V_\mathrm{SD} =$ \SI{-1}{\milli\volt} (we use the same convention for the bias sign as for the NW) is depicted in Fig. \ref{fig:Supp_Finfet}a where the reservoir transition shows segments similar to the data presented in Fig. \ref{fig:overviewfeature2}a. The corresponding segments are marked analogously, highlighting in the upper reservoir transition the $(1,1)\leftrightarrow(1,0)$ transition outside of the bias window (orange circle), the gap due to PSB lifting close to zero-detuning (orange triangle in the zoom-in in Fig. \ref{fig:Supp_Finfet}b), the bright PSB regime in the upper reservoir transition (orange star). In the lower reservoir transition, the PSB regime shows low visibility (green star) and the transition is recovered outside the bias window where the $(2,0)\leftrightarrow(2,1)$ (orange circle) reappears. Note the absence (compared to Fig. \ref{fig:overviewfeature2}a) of the high-visibility  feature close to zero detuning in the lower reservoir transition, presumably due to a different imbalance of the reservoir tunnel rates as compared to the NW DQD. Because the FinFET device only featured one dedicated interdot barrier gate which was set to $V_\mathrm{B} =$ \SI{-930}{\milli\volt} for all displayed experiments, the reservoir tunnel rates were only poorly tuneable and the impact of the reservoir tunnel rate imbalance could not be further studied.

An interesting side note is the visibility of the interdot transition along the triangle baselines. Due to a relatively large mutual capacitance, the triple points are separated in energy by a sufficient amount so that the triangles do not overlap for $V_\mathrm{SD} =$ \SI{-1}{\milli\volt}. The interdot transition is therefore partially present outside (light blue arrow) and inside (dark blue arrows) the bias window in this charge stability map. Evidently, the dynamics imposed by the bias voltage rearrange the population of the DQD states such that the triangle baselines disappear while the interdot transition outside of the bias window is very clearly visible.

Finally, the response of the DC current and the dispersive feature to a magnetic field is shown in Fig. \ref{fig:Supp_Finfet}c for a more negative bias $V_\mathrm{SD} =$ \SI{-3}{\milli\volt}. Unlike for the Ge/Si NW device, PSB in the FinFETs is usually lifted by magnetic fields only close to zero detuning \cite{Geyer2021,Camenzind2022}. For this reason, the current at $B_\mathrm{z} = $ \SI{0.5}{\tesla} is high only at the triangle baselines, as inelastic processes only weakly lift PSB, leading to a small increase in current towards the top left. This is also reflected in the dispersively sensed data where the upper reservoir transition largely prevails while the lower transition remains at low contrast. The slightly increasing gap in the upper transition close to zero detuning (purple circles) is in line with the observation of enhanced current. Note that the first excited state transition is not visible in these plots as it lies at energies higher than the applied bias.

Overall, this observation highlights the broad impact of our finding, as the dispersive signature is present across different device platforms and material systems.

\clearpage

%

\end{document}